# Robust Precoder for Mitigating Inter-Symbol and Inter-Carrier Interferences in Coherent Optical FBMC/OQAM

Khaled Abdulaziz Alaghbari*[1], Heng-Siong Lim[2], Tawfig Eltaif[3]

Faculty of Engineering & Technology, Multimedia University (MMU), Bukit Beruang, 75450, Melaka, *Malaysia*
*[1] Corresponding author email: *khalidazizxp@yahoo.com*,
[2] *hslim@mmu.edu.my*, [3] *tefosat@ieee.org*

This work was funded by Multimedia University (Malaysia), project SAP ID: MMUI/160105.

**Abstract:** In this paper, a new precoder for a coherent optical filter bank multicarrier with offset quadrature amplitude modulation (FBMC/OQAM) system is proposed. The precoder is designed based on an iterative polynomial eigenvalue decomposition (PEVD) algorithm to jointly mitigate the inter-symbol interference (ISI) and inter-carrier interference (ICI). The PEVD algorithm is used to decompose a polynomial channel matrix into polynomial eigenvectors and eigenvalues matrices, then a precoder is designed based on the decomposed matrices. The precoder acts as a filter applied to each subcarrier and to its adjacent subcarriers. In addition, a new adaptive optimum energy algorithm is proposed to truncate the insignificant precoder filter coefficients produced by the PEVD algorithm and consequently reduce the computational complexity. A mathematical model and algorithm for implementing the precoder are also presented. Robustness of the proposed precoder has been analyzed and compared to existing precoder in optical channel with different dispersion effects. Numerical results demonstrate that the new precoder significantly outperforms the existing precoder in terms of error vector magnitude (EVM), bit error rate (BER), Frobenius norm of the error matrix, and computational complexity.

## 1. Introduction

The Filter bank multicarrier with offset quadrature amplitude modulation (FBMC/OQAM) is an emerging waveform, which applies filter banks to the multicarrier signal and eliminates the need for cyclic prefix (CP) used in the popular orthogonal frequency division multiplexing (OFDM) system. The FBMC/OQAM system, which employs the Phydyas filter (a well-localized pulse-shaping filter in time and frequency domains), has recently attracted attention due to its spectrum efficiency, lower side-lobes, and robustness to narrow-band interference [1, 2]. Despite its advantages compared to OFDM, it generates a number of new signal processing challenges, which may not exist in other multicarrier modulation schemes [3]. The FBMC/OQAM system suffers from inherent intrinsic interference due to adjacent subcarriers when the Phydyas filter (or any other pulse shaping filter) is used [4, 5]. The problem is exacerbated by additional distortions found in the optical channel. Consequently, inter-symbol interference (ISI) and inter-subcarrier interference (ICI) on each subcarrier arise, making it necessary to apply advanced channel equalization. An alternative approach for mitigating these interferences is by using a precoder, which can be seen as a channel pre-equalizer at the transmitter. A precoder can significantly reduce the computational complexity needed at the receiver and consequently reduce the receiver's size and cost. There are some recent works that consider cancellation of intrinsic interference by coding matrix at the transmitter side, but at the receiver additional processing such as iterative interference cancellation and one-tap equalization [6] or de-coding [7] is still required.

In order to jointly mitigate the ICI and ISI, the space division multiple access (SDMA) technique normally adopted by wireless systems can be exploited to design the precoder. In [8], SDMA with FBMC/OQAM is combined with multiple-input multiple-output (MIMO) wireless systems. The authors in [8] investigated the system performance in high frequency selective wireless channels using a low number of subcarriers. The precoder is designed by inversion of composite channel matrix containing pulse shaping filter, channel distortion, and matched filter for each subcarrier. The results demonstrate improved performance compared to that of the canonical MIMO-OFDM system. However, a root raised cosine (RRC) pulse shaping filter is used, which is known to be less effective than the Phydyas filter proposed recently for the FBMC/OQAM system. Furthermore, inversion of the precoder channel matrix is a critical issue when dealing with a sparse matrix due to the singularity problem. Applying the Phydyas filter can significantly reduce the precoder channel matrix and consequently reduce the precoder computational complexity. In this paper, we propose a new precoder channel matrix and apply it to optical communication systems.

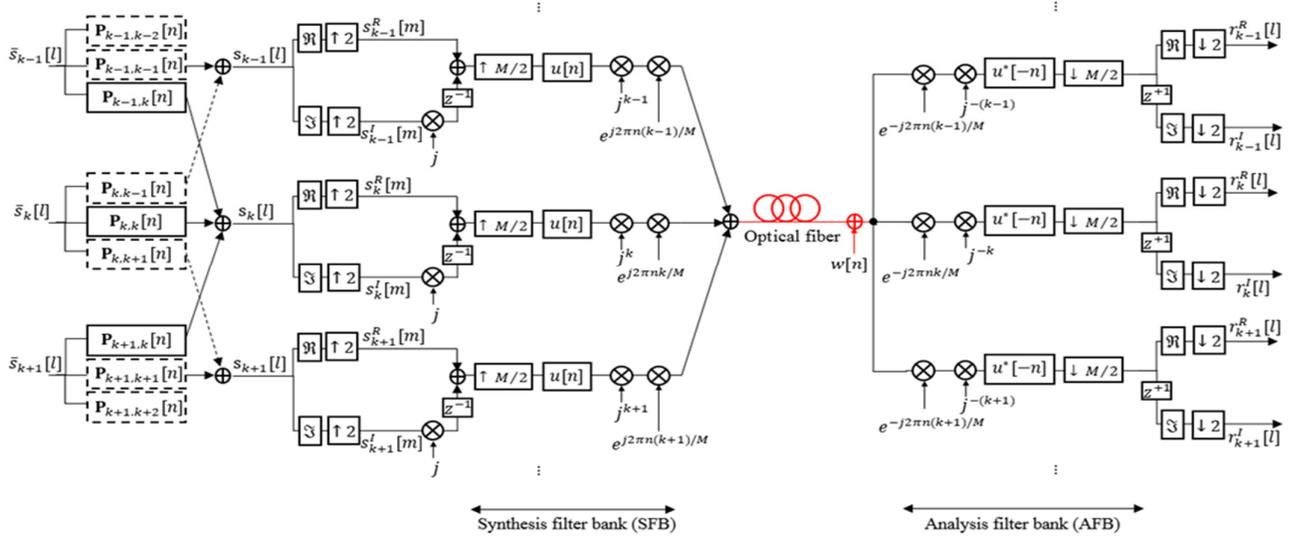

Fig. 1 FBMC/OQAM with precoder design for subcarrier *k* in optical channel

To invert the precoder channel matrix, the classical singular value decomposition (SVD) is no longer applicable since the precoder channel matrix is actually a polynomial matrix. An advanced technique such as the iterative polynomial eigenvalue decomposition (PEVD) can be used [9]. Two popular PEVD methods mostly used in this context are second-order sequential best rotation (SBR2) [10] and sequential matrix diagonalization (SMD) [11]. The SBR2 and SMD algorithms have been successfully applied to problems in broadband communication, spectral factorization, and blind source separation [12]. In [13], SBR2 has been parallelized and implemented using field programmable gate arrays to solve real-time problems. Many derived methods based on these two algorithms can be found in literature. Similar to SVD, the principal aim of the iterative PEVD is to decompose the input matrix into para-unitary and diagonal matrices. However, the generated polynomial matrices using PEVD techniques can be of high order, which limits their practicality. To reduce the computational complexity of the iterative PEVD algorithms, truncating small coefficients of the resulting matrices has been recommended in [14, 15]. In [14], truncation methods for para-Hermitian polynomial matrix had been evaluated for each iteration of the SBR2 algorithm and after convergence. This technique is based on discarding the coefficients with the smallest Frobenius norm. It is found that truncation of the para-unitary matrix after the diagonalization process yields lower matrix coefficients than that of the case when the trimming is done during the process of diagonalization. Moreover, truncation during the diagonalization process increases the error function, which has a negative impact on the algorithm performance. However, similar to [10], only a few matrix coefficients are trimmed. The implementation of the precoder may be extremely difficult if the precoder impulse response is excessively long [14, 15]. In this paper, we propose a new algorithm that is based on adaptive optimum energy, where only coefficients with significant energy will be selected for the precoder.

This paper is organized as follows: optical FBMC/OQAM system modelling is introduced in section 2; the mathematical derivations of the proposed precoder and its implementation algorithm are described in section 3; results and discussion are presented in section 4; finally, a conclusion is drawn in section 5.

## 2. System Model

Fig. 1 illustrates the optical FBMC/OQAM system with a precoder designed for subcarrier *k* to eliminate the ISI and ICI. The block diagram contains precoder block, conversion from QAM to offset-QAM, synthesis filter bank (SFB), optical channel, analysis filter bank (AFB), and finally conversion of OQAM to QAM symbols. The precoder is designed based on the inversion of SFB, the optical channel, and AFB.

### 2.1. FBMC/OQAM System Model

Let $s_k[l]$ represent the QAM symbol transmitted on subcarrier *k* at a symbol rate $R_s = 1/T_s$, where $T_s$ is the QAM symbol duration. As shown in Fig. 1, the QAM symbol is separated into its real and imaginary parts and up-sampled by 2 to obtain the real components $s_k^R[m]$ and imaginary components $s_k^I[m]$:

$$s_k^R[m] = \begin{cases} \Re\{s_k(l)\}, & m = 2l \\ 0, & \text{otherwise} \end{cases} \quad \text{and} \quad s_k^I[m] = \begin{cases} \Im\{s_k(l)\}, & m = 2l \\ 0, & \text{otherwise} \end{cases} \quad (1)$$

where $\Re\{.\}$ and $\Im\{.\}$ represent the real and imaginary parts, respectively. To implement the offset QAM signal $s_k[m]$, the sequence $s_k^I[m]$ is multiplied by $j = \sqrt{-1}$, delayed by half a symbol time, and then added with the sequence $s_k^R[m]$.

Next, the sequence is up-sampled by *M*/2 and the symbols are filtered with the Phydyas pulse shaping filter $u[n]$. The *M* subcarriers are summed up to obtain the transmitted signal [8]. The signal propagation inside the optical fiber is subjected to chromatic dispersion (CD) and additive white Gaussian noise $w[n]$ with variance $\sigma_w^2$. The CD effect can be represented in discrete time domain after applying proper truncation to avoid aliasing as follows [16]:

$$h[n] = \sqrt{\frac{jcT^2}{D\,\lambda^2\,L}} \exp\left(-j\,\frac{\pi c T^2}{D\,\lambda^2\,L}\,n^2\right), \text{where} -\left\lfloor\frac{N_{tap}}{2}\right\rfloor \leq n \leq \left\lfloor\frac{N_{tap}}{2}\right\rfloor \quad (2)$$

where *T* is sampling time, *L* is the fiber length, *D* is the dispersion coefficient, $\lambda$ is the carrier wavelength and *c* is the speed of light. The number of channel taps $N_{tap} = 2 \times \left\lfloor\frac{|D|\lambda^2 L}{2cT^2}\right\rfloor + 1$, and $\lfloor x \rfloor$ denotes the nearest integer less than $x$. At the receiver, an analysis filter bank (AFB) is used for demodulating the transmitted symbols. The received signal is passed through the filters matched to the Phydyas pulse shaping filters and the delay applied earlier at the transmitter is compensated so that both real and imaginary parts are recovered synchronously after down-sampling the signal by *M*. Only the real component on the real branch $r_k^R[l]$ and imaginary component on the imaginary branch $r_k^I[l]$ are kept at the receiver to reconstruct the QAM symbol $s_k[l]$. In the case of the frequency flat channel, the transmitted signal can be perfectly recovered at the receiver. However, as in most of the practical cases, the optical fiber channel is frequency selective and consequently the system suffers from ISI on each subcarrier and ICI between adjacent subcarriers. The objective of the precoder is to combat these optical channel distortions by pre-compensating the information symbols on a subcarrier basis so that the effects of the interferences are eliminated.

### 2.2. FBMC/OQAM Equivalent System Model

This section presents an equivalent system model of the system discussed in section 2.1 and Fig. 1. We define the two polyphase components of the transmitted sequences on subcarrier *i* as [8]:

$$s_{i,0}^R[l] := s_i^R[2l] \quad \text{and} \quad s_{i,1}^R[l] := s_i^R[2l+1] \quad (3)$$
$$s_{i,0}^I[l] := s_i^I[2l] \quad \text{and} \quad s_{i,1}^I[l] := s_i^I[2l+1] \quad (4)$$

where $i = -M/2, \ldots, M/2 - 1$. Based on the polyphase representations, the received sequences on subcarrier *k* can be written as:

$$r_k^R[l] = \sum_{i=-M/2}^{M/2-1} s_{i,0}^R[l] \otimes g_{i,k,0}^{RR}[l] + s_{i,1}^R[l] \otimes g_{i,k,1}^{RR}[l] + s_{i,0}^I[l] \otimes g_{i,k,0}^{IR}[l] + s_{i,1}^I[l] \otimes g_{i,k,1}^{IR}[l] + n_k^R[l] \quad (5)$$

$$r_k^I[l] = \sum_{i=-M/2}^{M/2-1} s_{i,0}^R[l] \otimes g_{i,k,0}^{RI}[l] + s_{i,1}^R[l] \otimes g_{i,k,1}^{RI}[l] + s_{i,0}^I[l] \otimes g_{i,k,0}^{II}[l] + s_{i,1}^I[l] \otimes g_{i,k,1}^{II}[l] + n_k^I[l] \quad (6)$$

where $\otimes$ is the convolution operator. The composite channel impulse responses, which are the results of the convolution of the transmitter filter (SFB), optical channel impulse response and receiver filter (AFB), are then down-sampled by a factor of *M*, so that the polyphase representations are given by:

$$\left.\begin{aligned}g_{i,k,0}^{RR}[l] &:= \Re\{p_i^R[n] \otimes h[n] \otimes q_k^R[n]\}_{|n=lM} \\ g_{i,k,1}^{RR}[l] &:= \Re\{p_i^R[n] \otimes h[n] \otimes q_k^R[n]\}_{|n=lM-M/2}\end{aligned}\right\} \text{ and } \left.\begin{aligned}g_{i,k,0}^{RI}[l] &:= \Im\{p_i^R[n] \otimes h[n] \otimes q_k^I[n]\}_{|n=lM} \\ g_{i,k,1}^{RI}[l] &:= \Im\{p_i^R[n] \otimes h[n] \otimes q_k^I[n]\}_{|n=lM-M/2}\end{aligned}\right\} \quad (7)$$

$$\left.\begin{aligned}g_{i,k,0}^{IR}[l] &:= \Re\{p_i^I[n] \otimes h[n] \otimes q_k^R[n]\}_{|n=lM} \\ g_{i,k,1}^{IR}[l] &:= \Re\{p_i^I[n] \otimes h[n] \otimes q_k^R[n]\}_{|n=lM-M/2}\end{aligned}\right\} \text{ and } \left.\begin{aligned}g_{i,k,0}^{II}[l] &:= \Im\{p_i^I[n] \otimes h[n] \otimes q_k^I[n]\}_{|n=lM} \\ g_{i,k,1}^{II}[l] &:= \Im\{p_i^I[n] \otimes h[n] \otimes q_k^I[n]\}_{|n=lM-M/2}\end{aligned}\right\} \quad (8)$$

where the SFB and AFB are also decomposed into two polyphase components and can be written as:

$$\left.\begin{aligned}p_i^R[n] &:= u[n]\,j^i \exp\left(j2\pi\frac{in}{M}\right) \\ p_i^I[n] &:= u[n-M/2]\,j^{i+1}(-1)^i \exp\left(j2\pi\frac{in}{M}\right)\end{aligned}\right\} \text{ and } \left.\begin{aligned}q_i^R[n] &:= u^*(-n)\,j^{-i} \exp\left(j2\pi\frac{in}{M}\right) \\ q_i^I[n] &:= u^*(-n-M/2)\,j^{-i}(-1)^i \exp\left(j2\pi\frac{in}{M}\right)\end{aligned}\right\} \quad (9)$$

The imaginary branches $p_i^I[n]$ and $q_i^I[n]$ are derived from the real branches such that their z-domain counterparts are $p_i^I(z) = j\,z^{-\frac{M}{2}}\,p_i^R(z)$ and $q_i^I(z) = z^{-\frac{M}{2}}\,q_i^R(z)$ respectively. The noise sequences $n_k^R[l]$ and $n_k^I[l]$ are obtained by filtering the input noise sequence $w[n]$ with the AFB and down-sampled by *M*:

$$n_k^R[l] = \Re\{w[n] \otimes q_k^R[n]\}_{|n=lM} \quad \text{and} \quad n_k^I[l] = \Im\{w[n] \otimes q_k^I[n]\}_{|n=lM} \quad (10)$$

The polyphase components of the input sequences can be arranged in a sequence of length-4 vectors $\mathbf{s}_i[l] :=$

$\begin{bmatrix} s_{i,0}^R[l] & s_{i,1}^R[l] & | & s_{i,0}^I[l] & s_{i,1}^I[l] \end{bmatrix}^T$. The noise and received sequences can be arranged in sequences of length-2 vectors $\mathbf{n}_k[l] := \begin{bmatrix} n_k^R[l] & | & n_k^I[l] \end{bmatrix}^T$ and $\mathbf{r}_k[l] := \begin{bmatrix} r_k^R[l] & | & r_k^I[l] \end{bmatrix}^T$ respectively. The composite channel impulse responses can also be arranged in a sequence of size-$2 \times 4$ matrices:

$$\mathbf{G}_{i,k}[l] := \begin{bmatrix} g_{i,k,0}^{RR}[l] & g_{i,k,1}^{RR}[l] & g_{i,k,0}^{IR}[l] & g_{i,k,1}^{IR}[l] \\ g_{i,k,0}^{RI}[l] & g_{i,k,1}^{RI}[l] & g_{i,k,0}^{II}[l] & g_{i,k,1}^{II}[l] \end{bmatrix} \quad (11)$$

Therefore, the model in Eq. (5) and Eq. (6) can be equivalently written as:

$$\mathbf{r}_k[l] = \sum_{i=-M/2}^{M/2-1} \mathbf{G}_{i,k}[l] \otimes \mathbf{s}_i[l] + \mathbf{n}_k[l] \quad (12)$$

where the convolution $\otimes$ of a vector sequence $\mathbf{x}[l]$ with a matrix sequence $\mathbf{G}[l]$ is defined as: $\mathbf{y}[l] := \sum_{m=-\infty}^{\infty} \mathbf{G}[m] \cdot \mathbf{x}[m-l]$. The expression in Eq. (12) can be written in the following form to explicitly indicate the problem of intrinsic interference:

$$\mathbf{r}_k[l] = \mathbf{G}_{k,k}[l] \otimes \mathbf{s}_k[l] + \underbrace{\sum_{\substack{j \in \Omega \\ j \neq k}} \mathbf{G}_{j,k}[l] \otimes \mathbf{s}_j[l]}_{\text{interference}} + \mathbf{n}_k[l] \quad (13)$$

where the set $\Omega$ includes all the subcarriers except $j = k$. The interfering channel matrix $\mathbf{G}_{j,k}[l]$ includes the TMUX response [4] and the optical channel impulse response as described in Eq. (7) and Eq. (8). A mathematical expression that involves all received subcarriers can be expressed in a unified matrix model as:

$$\begin{bmatrix} \vdots \\ \mathbf{r}_{k-2}[l] \\ \mathbf{r}_{k-1}[l] \\ \mathbf{r}_k[l] \\ \mathbf{r}_{k+1}[l] \\ \mathbf{r}_{k+2}[l] \\ \vdots \end{bmatrix} = \begin{bmatrix} & \vdots & \vdots & \vdots & \vdots & \vdots \\ \cdots & \mathbf{G}_{k-2,k-2}[l] & \mathbf{G}_{k-1,k-2}[l] & \mathbf{G}_{k,k-2}[l] & \mathbf{G}_{k+1,k-2}[l] & \mathbf{G}_{k+2,k-2}[l] & \cdots \\ & \mathbf{G}_{k-2,k-1}[l] & \mathbf{G}_{k-1,k-1}[l] & \mathbf{G}_{k,k-1}[l] & \mathbf{G}_{k+1,k-1}[l] & \mathbf{G}_{k+2,k-1}[l] \\ & \mathbf{G}_{k-2,k}[l] & \mathbf{G}_{k-1,k}[l] & \mathbf{G}_{k,k}[l] & \mathbf{G}_{k+1,k}[l] & \mathbf{G}_{k+2,k}[l] \\ & \mathbf{G}_{k-2,k+1}[l] & \mathbf{G}_{k-1,k+1}[l] & \mathbf{G}_{k,k+1}[l] & \mathbf{G}_{k+1,k+1}[l] & \mathbf{G}_{k+2,k+1}[l] \\ & \mathbf{G}_{k-2,k+2}[l] & \mathbf{G}_{k-1,k+2}[l] & \mathbf{G}_{k,k+2}[l] & \mathbf{G}_{k+1,k+2}[l] & \mathbf{G}_{k+2,k+2}[l] \\ & \vdots & \vdots & \vdots & \vdots & \vdots \end{bmatrix} \otimes \begin{bmatrix} \vdots \\ \mathbf{s}_{k-2}[l] \\ \mathbf{s}_{k-1}[l] \\ \mathbf{s}_k[l] \\ \mathbf{s}_{k+1}[l] \\ \mathbf{s}_{k+2}[l] \\ \vdots \end{bmatrix}$$
$$+ \begin{bmatrix} \vdots \\ \mathbf{n}_{k-2}[l] \\ \mathbf{n}_{k-1}[l] \\ \mathbf{n}_k[l] \\ \mathbf{n}_{k+1}[l] \\ \mathbf{n}_{k+2}[l] \\ \vdots \end{bmatrix} \quad (14)$$

$$\bar{\bar{\mathbf{r}}}[l] = \bar{\bar{\mathbf{G}}}[l] \otimes \bar{\bar{\mathbf{s}}}[l] + \bar{\bar{\mathbf{n}}}[l] \quad (15)$$

The unified model above shows the mutual impacts between all the subcarriers.

## 3. FBMC/OQAM Precoder Design

The precoder can be jointly designed for all subcarriers based on their mutual impact as revealed in the unified matrix model of Eq. (14). However, this option would be computationally expensive since it requires inversion of a very large polynomial matrix. It is possible to build a low complexity precoder based on the knowledge that each subcarrier only interferes with its two adjacent subcarriers [4, 17]. Therefore, the coefficients in the upper right triangular region and lower left triangular region in the channel matrix $\bar{\bar{\mathbf{G}}}[l]$ can be replaced with zeros since they have insignificant effects with symbols at subcarrier $k$. Therefore, we propose to inverse the following $6 \times 12$ channel matrix:

$$\mathbf{G}_k[l] = \begin{bmatrix} \mathbf{G}_{k-1,k-1}[l] & \mathbf{G}_{k,k-1}[l] & \mathbf{0}_{2\times 4} \\ \mathbf{G}_{k-1,k}[l] & \mathbf{G}_{k,k}[l] & \mathbf{G}_{k+1,k}[l] \\ \mathbf{0}_{2\times 4} & \mathbf{G}_{k,k+1}[l] & \mathbf{G}_{k+1,k+1}[l] \end{bmatrix} \quad (16)$$

The matrix $\mathbf{G}_k[l]$ includes most of the significant interferences (ISI and ICI) while it can be inverted efficiently since it avoids the singularity issues which exist in a sparse matrix (if a larger matrix is chosen) which will be discussed later. The precoder matrix $\mathbf{P}_k[l]$ of size $12 \times 2$ is then given by:

$$\mathbf{P}_k[l] := \begin{bmatrix} \mathbf{P}_{k,k-1}[l] \\ \mathbf{P}_{k,k}[l] \\ \mathbf{P}_{k,k+1}[l] \end{bmatrix} \quad (17)$$

The precoder is designed to fulfill two requirements: 1) there is no interference on subcarrier $k$ and; 2) the symbols on subcarrier $k$ and $k \pm 1$ do not mutually interfere with each other. The overall mathematical expression for the FBMC/OQAM system with precoding can then be written as:

$$\bar{\mathbf{r}}_k[l] = \mathbf{G}_k[l] \otimes \mathbf{P}_k[l] \otimes \bar{\mathbf{s}}_k[l] + \bar{\mathbf{n}}_k[l] \tag{18}$$

where $\bar{\mathbf{r}}_k[l] = [\mathbf{r}_{k-1}[l] \quad \mathbf{r}_k[l] \quad \mathbf{r}_{k+1}[l]]^T$ and $\bar{\mathbf{n}}_k[l] = [\mathbf{n}_{k-1}[l] \quad \mathbf{n}_k[l] \quad \mathbf{n}_{k+1}[l]]^T$ are the received signal and noise vectors on different subcarriers respectively and $\bar{\mathbf{s}}_k[l] = [\bar{s}_k^R[l] \quad \bar{s}_k^I[l]]^T$ is the real-value input symbol vector. In the z-domain, the precoder matrix can be obtained via the pseudo-inverse of the matrix $\mathbf{G}_k(z)$

$$\mathbf{P}_k(z) = \widetilde{\mathbf{G}}_k(z) \left(\mathbf{G}_k(z) \widetilde{\mathbf{G}}_k(z)\right)^{-1} \Theta \tag{19}$$

where $\mathbf{G}_k(z) = \sum_{l=-\infty}^{\infty} \mathbf{G}_k[l] z^{-l}$ is a polynomial matrix, $\widetilde{\mathbf{G}}_k(z)$ is the para-Hermitian of $\mathbf{G}_k(z)$ and $\Theta$ is composed of the middle two columns of an identity matrix $\mathbf{I}_6$ so that only the middle two columns of the resulting pseudo-inverse matrix are selected. Due to the polynomial channel matrix $\mathbf{G}_k(z)$, computing the inverse of the square matrix $\mathbf{G}_k(z) \widetilde{\mathbf{G}}_k(z)$ in Eq. (19) is not straightforward and an advanced technique such as PEVD is needed. Similar to SVD, the PEVD decomposes the matrix $\mathbf{G}_k(z) \widetilde{\mathbf{G}}_k(z)$ into polynomial eigenvalues and eigenvector matrices. Then, only inversion of the polynomial eigenvalues will be required. The accuracy of the precoder matrix obtained in this way depends on the structure of the channel matrix and the decomposed matrices. In addition, the resulting precoder matrix contains elements (filters) with very long polynomial (coefficients) and needs to be truncated to ensure its practicality. The next section in this paper is devoted to discussing the method to compute the precoder matrix from the channel matrix. To study the robustness of our proposed method, we have compared and analyzed its characteristics and the inversion process with conventional precoder.

### 3.1. Decomposition of Polynomial Channel Matrix

The principal aim of PEVD algorithms is to iteratively diagonalize a para-Hermitian matrix $\mathbf{R}(z)$ to obtain a para-unitary polynomial matrix $\mathbf{Q}(z)$ that contains the polynomial eigenvectors and a para-Hermitian diagonal polynomial matrix $\mathbf{A}(z)$ that contains the polynomial eigenvalues. Description of algorithms operation can be found in [10, 11]. In general, the SMD algorithm is known to transfer more amount of energy per iteration, thus it converges faster than SBR2, and it is computationally costlier than the SBR2. A matrix $\mathbf{R}(z)$ is called para-Hermitian if $\mathbf{R}(z) = \widetilde{\mathbf{R}}(z) \in \mathbb{C}^{M \times M}(z)$ where the para-Hermitian operator $\{\tilde{\cdot}\}$ Implies conjugate transpose and time reversal of the polynomials, i.e. $\widetilde{\mathbf{R}}(z) = \mathbf{R}^H(z^{-1})$. Mathematically the PEVD of a para-Hermitian matrix $\mathbf{R}(z)$ is given by:

$$\mathbf{R}(z) = \mathbf{Q}(z)\mathbf{A}(z)\widetilde{\mathbf{Q}}(z) \tag{20}$$

where $\mathbf{Q}(z)$ is a para-unitary polynomial eigenvectors matrix that fulfills the property $\mathbf{Q}(z)\widetilde{\mathbf{Q}}(z) = \widetilde{\mathbf{Q}}(z)\mathbf{Q}(z) = \mathbf{I}$, and $\mathbf{A}(z)$ is a diagonal polynomial eigenvalues matrix, where its on-diagonal elements $A_i(z)$ are arranged in descending order to achieve the spectral majorization property, i.e., their power spectral densities $A_i(e^{j\Omega})$ satisfy $A_i(e^{j\Omega}) \geq A_{i+1}(e^{j\Omega}), i = 1, \dots, M-1, \forall \Omega$ [9]. The spectral majorization property is the polynomial analogue to the way eigenvalues are ordered in the classical SVD technique [12]. Based on the PEVD, the inverse of $\mathbf{R}(z)$ is computed as $\mathbf{R}^{-1}(z) = \mathbf{Q}(z)\mathbf{A}^{-1}(z)\widetilde{\mathbf{Q}}(z)$ where the inverse matrix $\mathbf{A}^{-1}(z)$ can be obtained by inverting the elements on the main diagonal of $\mathbf{A}(z)$. A technique called time domain inversion method presented in [9] can be used to find the inverse. This technique is based on constructing a convolution matrix for the polynomial eigenvalues $a_i[\tau]$ and then compute its inverse.

The steps required to implement the proposed precoder for the *k*-th subcarrier can be summarized as follows:

*Begin*
 *for i = k-1:k+1*
  1) *Calculate the para-Hermitian matrix $\mathbf{R}_i(z)$ from the channel matrix $\mathbf{G}_i(z)$,*
$$\mathbf{R}_i(z) = \mathbf{G}_i(z)\widetilde{\mathbf{G}}_i(z) \tag{21}$$
  2) *Use PEVD to decompose $\mathbf{R}_i(z)$ into $\mathbf{Q}_i(z)$ and $\mathbf{A}_i(z)$*
  3) *Use the time domain inversion technique to inverse $\mathbf{A}_i(z)$*
  4) *Calculate the inverse of the channel matrix as follows*:
$$\mathbf{G}_i^{-1}(z) = \widetilde{\mathbf{G}}_i(z) \left(\mathbf{Q}_i(z)\mathbf{A}_i^{-1}(z)\widetilde{\mathbf{Q}}_i(z)\right) \tag{22}$$
  5) *The precoder matrix of size $12 \times 2$ can be obtained as*:
$$\mathbf{P}_i(z) = \mathbf{G}_i^{-1}(z) \, \Theta \tag{23}$$
   *where $\Theta$ is a matrix consists of the two central columns of identity matrix $\mathbf{I}_6$.*
  6) *Precode the source signal to obtain a matrix of size $12 \times 1$*
$$\bar{\mathbf{P}}_{i,k}(z) = \mathbf{P}_i(z)\bar{\mathbf{s}}_i(z) \tag{24}$$
 *end*
  7) *Compute the precoded signal as:* $\mathbf{s}_k(z) = \sum_{i=k-1}^{k+1} \bar{\mathbf{P}}_{i,k}(z)$ (25)
*end*

### 3.2. Truncation of the Precoder Filters

The number of iterations for both SBR2 and SMD can be set in advance. With $N$ iterations, the output polynomial matrices will gradually increase in order. Increasing the order of output matrices will increase the precoder complexity and this limits its practicality. Another parameter, $\mu$, is available in SBR2 and SMD algorithms to truncate the output matrices and control the growth in the order of output matrices [10, 11]. However, this truncation method can decrease the efficiency and robustness of our precoder; thus, we proposed a new technique to curtail the precoder polynomial matrix. The proposed technique is based on adaptive optimum energy criteria, where only the significant precoder taps with optimum energy will be selected for precoding the source signal. The algorithm is based on the ratio of precoder energy considering $L$ taps to that of considering $L+1$ taps, where $L$ is first initialized and iteratively updated. Specifically, this ratio can be calculated as follows:

$$\sum_{n=0}^{L-1}|p_{k,ij}[n]|^2 \bigg/ \sum_{n=0}^{L}|p_{k,ij}[n]|^2 \qquad (26)$$

where $p_{k,ij}[n]$ is the $(i,j)$th element of the precoder polynomial matrix $\mathbf{P}_k(z)$.

The proposed algorithm can be summarized as follows:

*Begin*
   *Obtain the precoder matrix $\mathbf{P}_k(z)$ according to (23);*
   *Initialization: $\alpha, \beta, L_A = 1, L_B = 1$;*
   *for each element (filter) of $\mathbf{P}_k(z)$*
      $max = \arg \max_n (p_{k,ij}[n])$;
      *Compute ratio* $A = \sum_{n=max}^{max+L_A-1}|p_{k,ij}[n]|^2 / \sum_{n=max}^{max+L_A}|p_{k,ij}[n]|^2$;
      *while* $(A \leq \alpha)$
         $L_A = L_A + 1$;
      *end*
      *Compute ratio* $B = \sum_{n=max}^{max-L_B+1}|p_{k,ij}[n]|^2 / \sum_{n=max}^{max-L_B}|p_{k,ij}[n]|^2$;
      *while* $(B \leq \beta)$
         $L_B = L_B + 1$;
      *end*
      *Obtain the element of truncated precoder;*
   *end*
   *Construct the new truncated precoder matrix $\mathbf{P}_{k,\text{truncated}}(z)$;*
*end*

## 4. Result and Discussion

Unless otherwise stated, the following parameters were assumed in the simulations: number of subcarriers $M = 64$, 30Gbaud FBMC/OQAM employing 4-QAM, Phydyas filter with overlapping factor $K = 4$, laser source with carrier wavelength $\lambda = 1550$nm, 80km standard single mode fiber (SSMF) with dispersion of 17ps/nm/km, and 30 iterations for PEVD algorithms.

### 4.1. Precoder Analysis

Fig. 2 shows the polynomial eigenvalues matrix of the conventional precoder (Fig. 2a) [8] and the new precoder (Fig. 2b) after applying the SBR2 algorithm. It can be seen from the figures that off-diagonal elements have been eliminated by the PEVD, and only on-diagonal elements remain. The polynomial length of the eigenvalues in Fig. 2a and 2b are 147 and 173 respectively. For the eigenvalues of the conventional precoder, there is a significant amplitude drop after the sixth on-diagonal elements, which indicate that the $R_k(z)$ considered by the conventional precoder is ill conditioned and this will seriously impact the precoder performance. On the other hand, as shown in Fig. 2b, by reducing the size and zeros in the channel matrix, the eigenvalue matrix obtained in the proposed technique achieves full rank.

In Fig. 3 we investigated the two important properties of the resulting eigenvalues. The polynomial eigenvalues matrix $\mathbf{A}_k(z)$ is para-Hermitian and diagonal where its on-diagonal elements $a_i[\tau]$ are symmetrical (as in Fig. 3a) and any deviation from this property must be due to numerical issues during the inversion process of the eigenvalues [9]. In addition, the power spectral densities $A_i(e^{j\Omega})$ are in descending order as shown in Fig. 3b which indicates that the SBR2 algorithm has effectively converged to produce spectrally majorized para-Hermitian eigenvalues. From Fig. 3, it is clear that the proposed method satisfies both the symmetry and spectral majorization requirements. After applying the time domain inversion method [9], the inverted eigenvalue must also preserve the symmetrical property. Fig. 3c shows the power spectral densities of the inverse of the eigenvalues matrix in Fig. 3b. The time domain inversion approach shows good accuracy for inversion. The magnitude of the eigenvalue is positive (dB) values and its inverse is in the corresponding negative values; also, the curves are symmetric over the *x* axis.

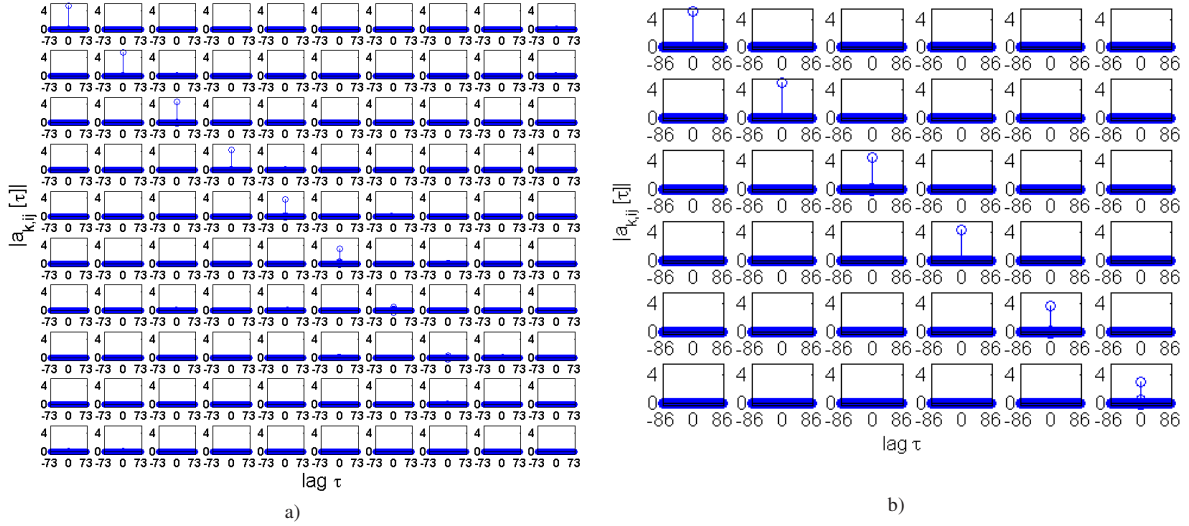

Fig. 2 Example of polynomial eigenvalues matrix $\mathbf{A}_k(z)$ produced by SBR2 for a) conventional b) proposed precoder

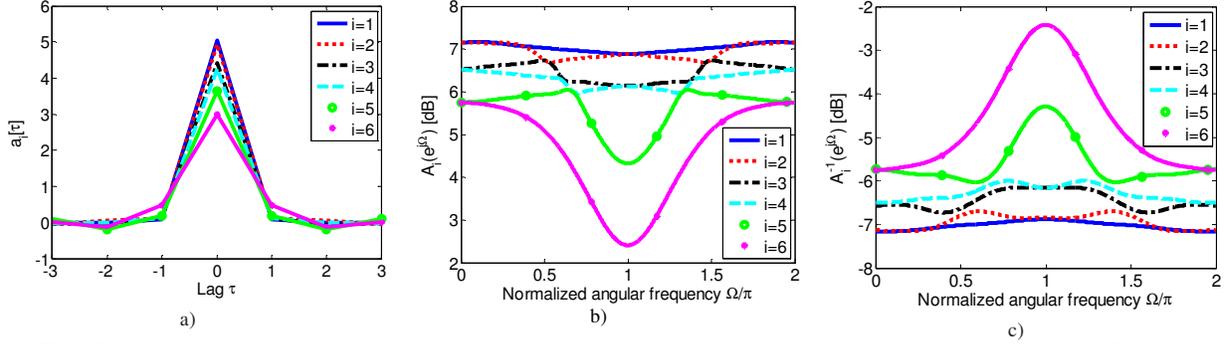

Fig. 3 Properties of the polynomial eigenvalues in a) time-lag b) spectral c) their inverse for the proposed precoder example given in Fig. 2b

The delay parameter in the time domain inversion method [9] is a crucial parameter. It should be optimized in order to achieve best performance of the technique. To this end, the error vector magnitude (EVM) [18] can be used as the performance metric. Fig. 4a shows that the optimum delay is 5 or above using either SBR2 or SMD for the proposed precoder. For the conventional precoder, the optimum delay for SBR2 also should be greater than 5; however, for SMD the EVM becomes worse when the delay is increased. This is due to the sparse structure of the conventional matrix as will be discussed further later. In Fig. 4b we have used SBR2 in the proposed precoder and tested different delay ranges when different subcarrier numbers were used. By using different subcarrier numbers, the delay should be also greater than 5 to get the best performance. This also means increasing the subcarrier number has no effect on the best delay used in the inversion process. Fig. 4b also shows that increasing the number of subcarriers will decrease the EVM. This is due to the fact that when using higher number of subcarriers, the symbol duration is longer and channel effect becomes flat.

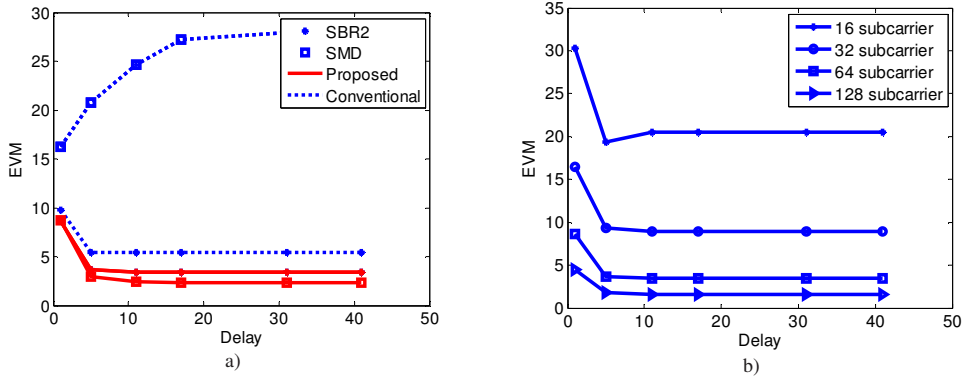

Fig. 4 Optimum delay using a) 64 subcarriers with SBR2 and SMD b) different number of subcarriers for proposed precoder with SBR2

In order to quantify how much the matrix $\mathbf{G}_k^{-1}(z)\mathbf{G}_k(z)$ deviates from the identity matrix $\mathbf{I}(z)$ due to decomposition and inversion of eigenvalues, we consider the Frobenius norm error given by [9, 14]:

$$\chi_k = \|\mathbf{I}_k(z) - \mathbf{G}_k^{-1}(z)\mathbf{G}_k(z)\|_F^2 \qquad (27)$$

where the Frobenius norm of a polynomial matrix $\mathbf{G}(z) = \sum_{v=-\infty}^{\infty} \mathbf{G}_v z^{-v}$ is defined as $\|\mathbf{G}(z)\|_F = \left(\sum_{v=-\infty}^{\infty} \|\mathbf{G}_v\|_F^2\right)^{1/2}$. As shown in Fig. 5b, the proposed precoder matrix gives strong identity polynomial matrix where the off-diagonal elements are completely eliminated compared to conventional precoder matrix shown in Fig. 5a. The remaining off-diagonal elements occur due to the sparse structure of the conventional precoder matrix, which causes an undesirable impact on the performance of the conventional precoder.

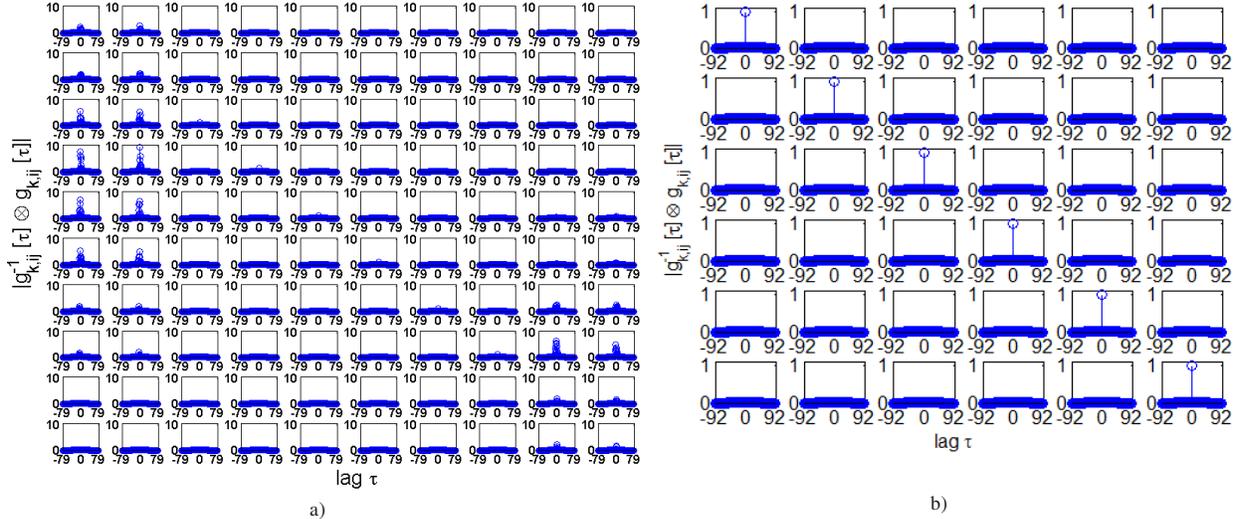

Fig. 5 $|\mathbf{G}_k^{-1}(z)\mathbf{G}_k(z)|$ of a) conventional and b) proposed precoder for the example given in Fig. 2

Fig. 6 shows the Frobenius norm error (in dB) for the proposed and conventional precoder matrix using 32 subcarriers after 80km of fiber length. Different subcarriers will encounter different channel distortions. Using the proposed precoder based on SBR2 or SMD, the Frobenius norm error is negative and slightly lower for SMD compared to that of SBR2, as presented in Fig. 6a. However, due to $|\mathbf{G}^{-1}(z)\mathbf{G}(z)|$ deviates very much from an identity matrix in conventional precoder as shown in Fig. 5a, the Frobenius norm error is positive and the difference in performance of SBR2 and SMD is obvious in Fig. 6b. The inaccurate inverse eigenvalues in the conventional precoder have a major impact on the precoder performance. Since the same optical channel realization is considered for both the SMD and SBR2 algorithms, the two curves appear to be correlated (especially in Fig. 6a). The resulting Frobenius norm error using SMD is higher compared to that of SBR2 and this explains the degraded performance of SMD in the conventional precoder matrix.

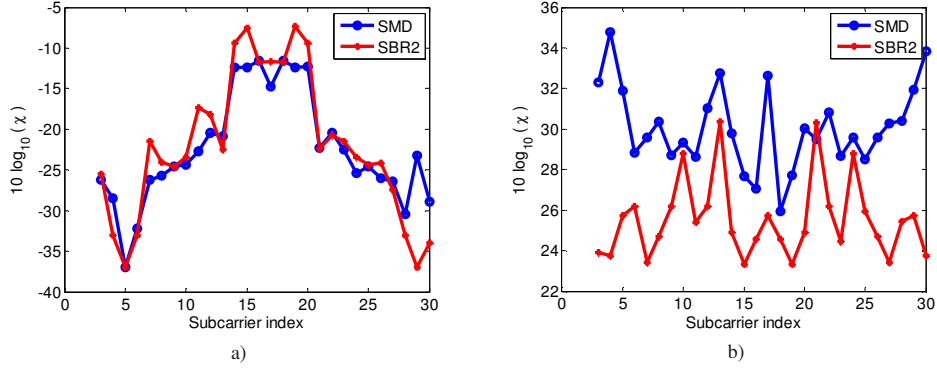

Fig. 6 Frobenius norm error of the a) proposed and b) conventional precoder

In Fig. 7 we compare the performance of the proposed precoder and the conventional precoder in terms of bit error rate (BER) versus signal to noise ratio (SNR). Both SBR2 and SMD algorithms are considered where both algorithms use 30 iterations. A 30G baud FBMC/OQAM with 64 subcarriers is transmitted over an optical transmission fiber of 180km. The BER curve for a back-to-back connection of transmitter and receiver scenario is also shown in the same figure as a reference for comparison. As shown in Fig. 7, the proposed precoder outperforms the conventional precoder using SBR2 or SMD. The proposed precoder is able to get rid of most of the inter-symbol and inter-subcarrier interference more effectively than the conventional precoder. For the proposed precoder, SMD outperform SBR2 when SNR > 8 dB, and improvement of >1dB can be achieved at higher SNR. The conventional precoder considers the inter-subcarrier interference from $k$-2 to $k$+2, which is unnecessary when employing the Phydyas filter. Also, due to the weakness in the structure of the precoder matrix, the SBR2 and SMD algorithms show degraded BER performance compared to the proposed precoder.

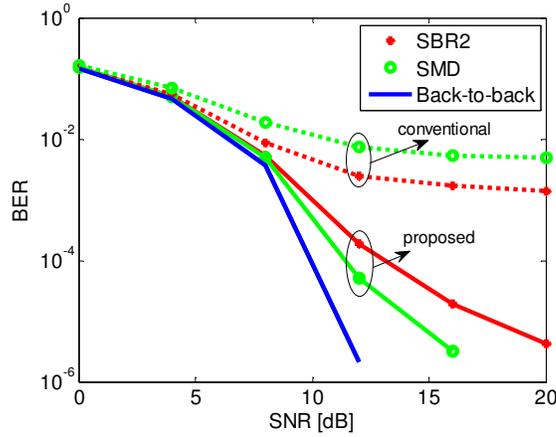

Fig. 7 BER versus SNR for 30G baud FBMC/OQAM after 180km transmission distance

Fig. 8a illustrates the performance of the conventional and new precoder over different transmission distances, and in 30 Gbaud and 20 Gbaud scenarios. To obtain this result, the SNR is fixed at 12 dB and the SBR2 algorithm with 30 iterations is used for both systems to have fair comparisons. The figure shows that EVM is gradually increasing as the transmission distance increases. In fact, the chromatic dispersion effect grows as the fiber gets longer. Performance of the FBMC/OQAM at 30 Gbaud degrades further due to a higher dispersion effect in the fiber. The results also demonstrate that the proposed system has better performance and tolerance to chromatic dispersion, while the performance of the conventional precoder is much worse especially with higher chromatic dispersion in the fiber. The figure also shows the diffusion of 4QAM constellation points caused by the channel at 80km and 400km.

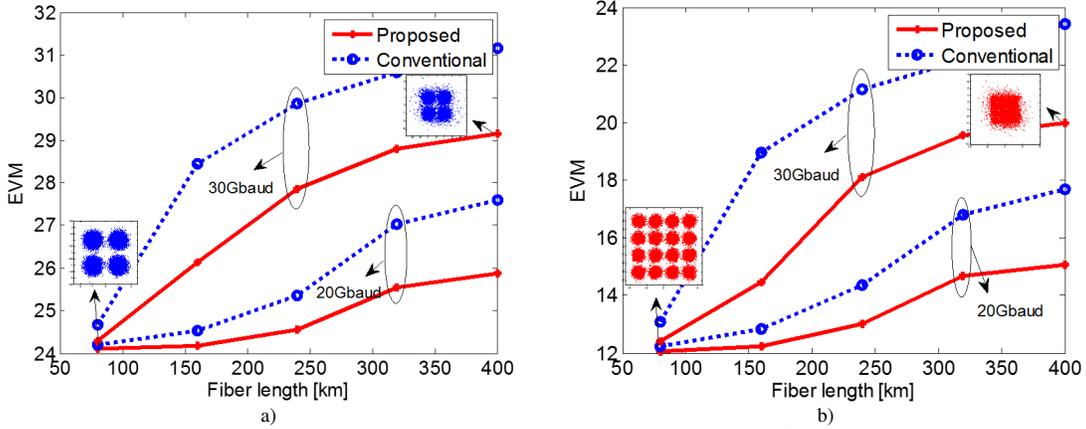

Fig. 8 EVM versus fiber length in a) FBMC-4OQAM at SNR = 12dB b) FBMC-16OQAM at SNR = 18dB

To highlight the robustness of the proposed precoder, we evaluated our system at a higher modulation order. Fig. 8b shows the dependence of EVM on the fiber length for FBMC/OQAM employing 16 QAM in 20G and 30G baud rate scenarios. The results are obtained at an SNR of 18 dB. Chromatic dispersion has greater impact on the system performance as shown in the constellation diagrams. At 80 km, the EVM performances of both precoders are between 12%- 13%. By increasing the fiber length to >100km, the new precoder again achieved the best performance in the two scenarios.

### 4.2. Precoder Computation Complexity

The computation complexity of the proposed and conventional precoders involves mainly the decomposition of the matrix $\mathbf{R}_k(z)$ into two matrices (polynomial eigenvalues and eigenvectors matrices) by using the PEVD algorithm, and then inversion of the polynomial eigenvalues matrix using the time domain inversion method. The number of subcarriers and number of PEVD algorithm iterations are the two important parameters that need to be addressed since they have a major impact on the computational complexity of the precoder. Fig. 9a shows the computational time computed by the Matlab software over different subcarrier numbers. The SBR2 algorithm with $N = 30$ iterations is considered, and the delay time required for the inversion process is assumed to be 11 for both precoders. As shown in Fig. 9a, the computational time increases linearly with the number of subcarriers. The proposed precoder achieved significantly lower time, especially when the number of subcarriers is more than 32. The computational complexity reduction realized by the proposed method is due to the smaller size matrices used in the decomposition and inversion algorithms. In Fig. 9b, we assessed the computational time that is required when different iterations of $N$ for the SBR2 algorithm are used. The computational time is averaged over 64 subcarriers; in other words, the result shown is only for one subcarrier. The computational time increases with the number of iterations $N$. The proposed precoder can achieve improvement of more than 0.2 second over different iteration numbers.

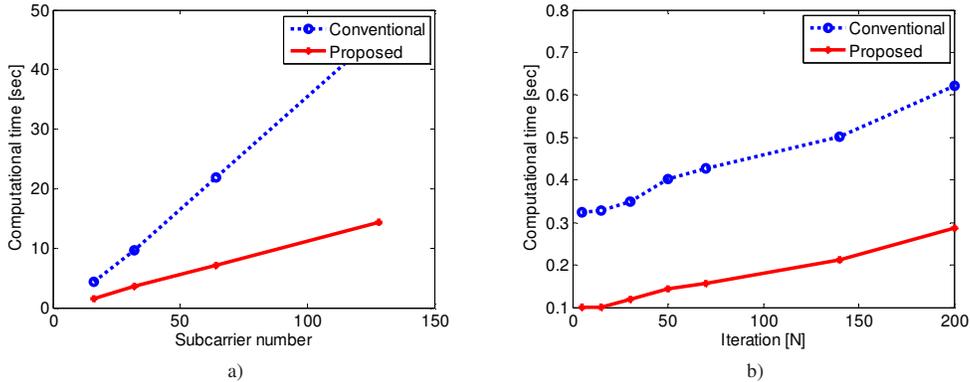

Fig. 9 Computation time of the proposed and conventional precoders using SBR2 algorithm versus a) subcarrier number b) iteration

### 4.3. Precoder Truncation

In Fig. 10, we compare the performance of SBR2 and SMD when different numbers of iterations is used. It shows that the proposed precoder matrix provides lower EVM compared to conventional precoder. This is the advantage of the proposed precoder, where a small number of iterations is needed. A higher number of iterations will result in higher computational complexity. At a low number of iterations, there is >1% difference in EVM between SBR2 and SMD, whereas at a higher number of iterations, both algorithms converge to about the same EVM. For the conventional precoder matrix, there is a noticeable performance difference between SBR2 and SMD due to the sparse structure of the conventional precoder matrix.

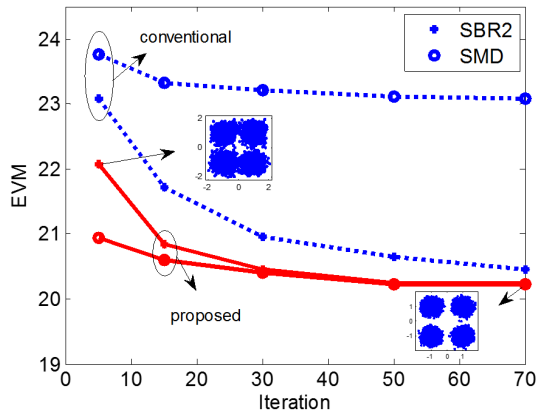

Fig. 10 EVM versus iterations in 30GBaud, fiber length = 80km and SNR = 10dB

In order to quantify how close the truncated channel matrix $\mathbf{G}_T^{-1}(z)$ is to $\mathbf{G}^{-1}(z)$, we evaluate how much the matrix $\mathbf{G}_T^{-1}(z)\,\mathbf{G}(z)$ deviates from an identity matrix $\mathbf{I}(z)$ due to truncation by using the Frobenius norm error given earlier in Eq. (27). As shown in Fig. 11a, without truncation, the length of the precoder matrix is high due to the high order of a para-unitary matrix produced by the SBR2 algorithm. It can be reduced after the diagonalization process of SBR2 when we set $\mu = 10^{-3}$, but the deviation from the identity matrix error will increase. If we further decrease $\mu = 10^{-5}$, the error will decrease but the precoder matrix length will remain high as illustrated in Fig. 11b. Using the proposed technique with $\alpha = \beta = 0.9$, the length of the precoder matrix is significantly reduced with almost the same deviation error as the no truncation case. In other word, the proposed technique aims to keep the loss in the precoder as low as possible with significant reduction in the polynomial order.

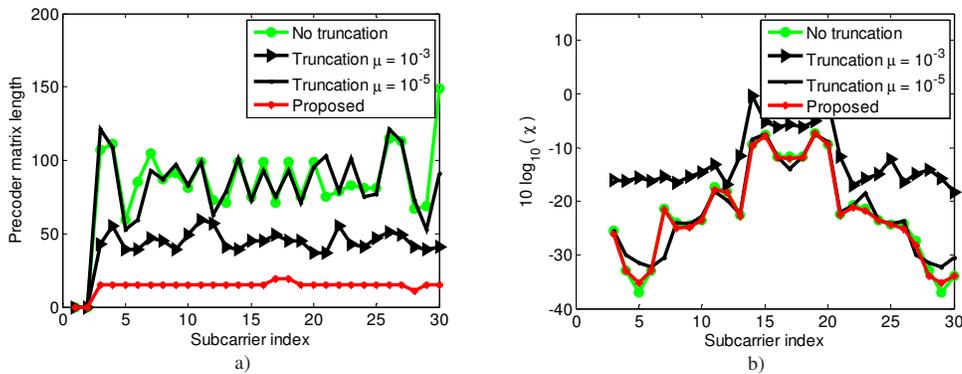

Fig. 11 a) Precoder matrix length after truncation b) corresponding Frobenius norm error using SBR2 with 30 iterations in 80km of fiber length, SNR = 10dB

Fig. 12 depicts the truncation impact on BER performance as a function of SNR. The result for back-to-back connection is plotted as a benchmark. Also, the result for no truncation is plotted as a reference for comparison and to show how various truncation techniques deviate from the reference curves. The proposed technique shows good and close performance to reference curve. It discards the outer coefficients, which have smaller or insignificant energy, by using the proposed adaptive optimum energy algorithm. The result for truncation with $\mu = 10^{-5}$ deviates slightly from the no truncation case. However, the precoder matrix length is still higher as discussed earlier. The truncation with $\mu = 10^{-3}$ causes degradation in the system performance. Increasing the fiber length to 400 km will increase the chromatic dispersion and yield degradation in the system performance. No obvious difference can be seen for the cases of no truncation, optimum energy technique (proposed), and truncation with $\mu = 10^{-5}$. However, truncation with $\mu = 10^{-3}$ results in serious deterioration of the system performance at 400km.

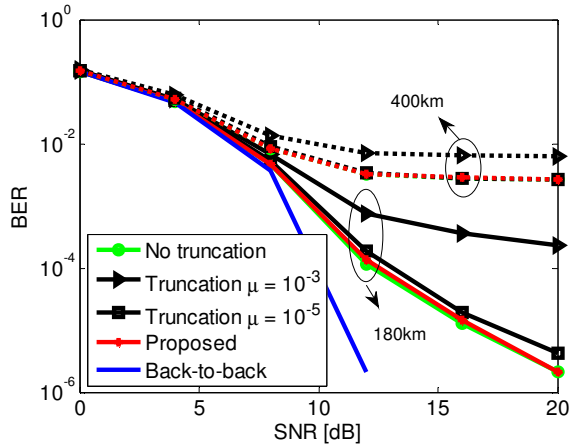

Fig. 12 BER versus SNR for precoded optical FBMC/OQAM system

## 5. Conclusions

In this paper, we have proposed a new precoder for mitigating inter-symbol and inter-carrier interferences by exploiting the mutual impact between the subcarriers in optical FBMC/OQAM systems. We have presented comparative analysis of the proposed precoder and the conventional precoder. Different performance criteria such as error vector magnitude (EVM), bit error rate (BER), Frobenius norm error, and computational complexity are used to quantify and measure the robustness and advantages of the proposed precoder. The derived precoder has been numerically validated in the optical FBMC/OQAM system with 4- and 16-QAM modulations, confirming its effectiveness. In addition, a new adaptive optimum energy algorithm is proposed to truncate the insignificant precoder filter coefficients and consequently reduce the computational complexity of the proposed precoder.